# Bacteria optimize tumble bias to strategically navigate surface constraints

Antai Tao[1], Guangzhe Liu[1,2,3], Rongjing Zhang[1,*], Junhua Yuan[1,*]

[1] Hefei National Research Center for Physical Sciences at the Microscale and Department of Physics, University of Science and Technology of China, Hefei, Anhui 230026, China.
[2] Wenzhou Institute, University of Chinese Academy of Science, Wenzhou, Zhejiang 325000, P.R. China
[3] School of Engineering and science, University of Chinese Academy of Science, Beijing 100049, P.R. China

Corresponding authors:
Junhua Yuan or Rongjing Zhang, Hefei National Research Center for Physical Sciences at the Microscale, and Department of Physics, University of Science and Technology of China, Hefei, Anhui, 230026, China, tel +86-551-63600851, email: jhyuan@ustc.edu.cn; rjzhang@ustc.edu.cn

## Abstract

In natural environments, solid surfaces present both opportunities and challenges for bacteria. On one hand, they serve as platforms for biofilm formation, crucial for bacterial colonization and resilience in harsh conditions. On the other hand, surfaces can entrap bacteria for extended periods and force them to swim along circular trajectories, constraining their environmental exploration compared to the freedom they experience in the bulk liquid. Here, through systematic single-cell behavioral measurements, phenomenological modeling, and theoretical analysis, we reveal how bacteria strategically navigate these factors. We observe that bacterial surface residence time decreases sharply with increasing tumble bias from zero, transitioning to a plateau at the mean tumble bias of wild-type *Escherichia coli* (~ 0.25). Furthermore, we find that bacterial surface diffusivity peaks near this mean tumble bias. Considering the phenotypic variation in bacterial tumble bias, which is primarily induced by noise in gene expression, this reflects a strategy for bacterial offspring persistence: In the absence of stimulus cues, some bacteria swiftly escape from the nearby surface in case it lacks nutrients, while others, with longer surface residence times, explore this two-dimensional environment most efficiently to find potential livable sites.

## 1. Introduction

Bacteria frequently encounter various surfaces in their natural habitats or in vivo. Solid surfaces serve as platforms for bacteria to form biofilms, which constitute the primary mode of bacterial growth in nature and enhance their resilience to environmental challenges[1,2]. The life cycle of bacterial biofilm is initiated following surface contact by planktonic cells, and followed by reversible attachment, irreversible attachment, biofilm maturation and final dispersion[3]. Moreover, these surfaces also significantly influence planktonic bacteria, causing them to accumulate at solid surfaces through a complex interplay of mechanisms, including hydrodynamic interaction, Brownian motion, and steric collision[4-9]. A representative example of self-propelled microorganism motion is planktonic *Escherichia coli* (*E. coli*), which exhibits the classic pattern of peritrichous bacterial motion[10]. When all the flagella rotate counterclockwise (CCW), they form a bundle, and the cell swims smoothly in a 'run' state. When some flagella rotate clockwise (CW), the flagellar bundle is disrupted, leading to a change in the swimming direction, known as the 'tumble' state[10,11]. As a swimming *E. coli* cell approaches a solid surface, it experiences surface-induced hydrodynamic forces that guide it to swim with its front pole inclined toward the surface[9], whereas electrostatic repulsion and steric forces at the point of contact tend to reorient the cell body parallel to the surface[8,12,13]. Consequently, the cell achieves an average, stable 'nose-down' configuration at equilibrium[8]. Despite the suppression of abrupt direction changes near the surface[7], tumble events remain the dominant mechanism for cells to escape from surface entrapment[14]. These physical mechanisms result in inefficient escape from the surface, leading to the confinement of bacterial motion in two dimensions on the surface plane for an extended period. Furthermore, hydrodynamic interactions of the counter-rotating cell body and flagellar bundle with the surface induce a torque on the cell, causing the running cell to move in a clockwise direction (when viewed from above), resulting in circular trajectories near surface[4,12,15]. This strongly affects the efficiency of surface exploration and raises the question of how *E. coli* cells modulate motion behaviors under these physical constraints of their habitat.

*E. coli* cells possess a chemotaxis signaling pathway that regulates their run-and-tumble behaviors to move toward a favorable environment[16]. However, bacteria often swim in natural environments where stimuli gradients are too weak to be detected as effective chemotaxis cues. Under these conditions, the fraction of time during which a single flagellar motor spins clockwise (referred to as CW bias) in wild-type *E. coli* cells is normally distributed within a narrow range (0-0.3 for most cells), with a mean value of ~ 0.12[17,18]. It has been demonstrated that a single clockwise-rotating flagellum, which breaks free from the flagellar bundle, is sufficient to induce a tumble event, namely the 'veto model'[17]. Consequently, the tumble bias (TB), which denotes the fraction of time that a bacterium spends in the tumble state and is closely related to the motor CW bias, should also exhibit a unimodal (approximately Gaussian) distribution for wild-type *E. coli* cells (Supplementary Note S1)[17]. This suggests that bacteria seem to adjust their run-tumble ratio to a narrow range in the absence of stimulus cues, which we measured in the present study. The significance of this adjustment is not yet understood, because the experimentally measured wild-type *E. coli*'s tumble bias and frequency do not maximize bacterial three-dimensional diffusivity in the bulk liquid[10].

Considering that pusher-type bacteria, whose rear-mounted flagella push the cell body forward, tend to accumulate near solid surfaces[4-8,19], and the primary mechanism by which *E. coli* cells escape from those surfaces has been proven to be tumble behaviors[14], here, we speculate the unimodal distribution of TB of *E. coli* cells in uniform environments should be closely relevant to their near-surface motion. To understand the significance of this distribution, we quantified the run-and-tumble behaviors and the duration *E. coli* cells spend on surfaces (surface residence time) using dual-optical trapping combined with a novel three-dimensional tracking technique. We observed a sharp decrease in surface residence time with increased TB, which then plateaued. The transition point between the decreasing and stable regions corresponds closely to the mean TB of wild-type *E. coli* cells. We further proposed a phenomenological model to approximate the bacterial near-surface motion, which reproduced the experimentally measured relationship between surface residence time and TB. Employing this model, we explored how bacterial diffusivity near surfaces varied with different TB by simulation. Interestingly, we found that during their surface residence periods, *E. coli* cells with TB near the mean value of the wild-type population exhibit maximum surface diffusivity. This finding was further confirmed by theoretical analysis, revealing an optimal exploration strategy for *E. coli* near surfaces.

## 2. Results

### 2.1 Relationship between the surface residence time and tumble bias for individual *E. coli* cells

*E. coli* strain HCB1736 ($\Delta cheY$), transformed with the plasmid pBAD33-*cheY*$^{13DK106YW}$, was used for the experiments. The expression level of CheY$^{13DK106YW}$ was induced to cover the entire range of run-and-tumble behaviors, encompassing TB from 0 to 1. For the measurement of run-and-tumble behaviors, a randomly selected cell with excellent motility was trapped by two separated optical tweezers for 180 s to record the motion of the two cell poles (Figure 1a and Movie S1). The instantaneous body wobble frequency ($\Omega_b$) can be derived and used to differentiate between run (high frequency) and tumble (low frequency) periods (Figure 1b and Experimental Section)[20]. We further compared the average $\Omega_b$ during the initial period (0-10 s) with that at the end of the trapping process (170-180 s). The average $\Omega_b$ ratio ($n=20$ cells) of the latter to the former was $1.01 \pm 0.11$ (Mean ± SD), indicating that bacterial proton motive force and viability were not notably affected by the trapping process. Furthermore, the average length of trapped *E. coli* cells was $2.4 \pm 0.4$ μm (Mean ± SD, distribution shown in Figure S1), in good agreement with previous measurements[21], confirming that the random selected bacterial samples represented the majority of the population. Nevertheless, we observed that very long cells, which comprised only a small fraction of the population, commonly exhibited relatively poor motility and were therefore excluded from the samples.

After measuring run-and-tumble behaviors, the same trapped cell was brought near the surface and released from the optical tweezers, and the following bacterial motion near the surface was recorded using bright-field microscopy. The definition of surface residence time was similar to that used in a recent study[14]. As illustrated in Figure 1c, two height criteria were employed to identify when a bacterium arrived at and escaped from the surface. The z-coordinate of the cell centroid ($z_{cell}$) was set to 0 when the cell body made contact with the surface and the cell axis was parallel to the surface plane. A bacterium was considered to be in the bulk region when $z_{cell} > 8$ μm. Effective interaction between the bacterium and the surface was considered to occur when $z_{cell} < 3$ μm. A surface arrival-escape event was defined as the process: $z_{cell} > 8$ μm → $z_{cell} < 3$ μm → $z_{cell} > 8$ μm. The surface residence time ($T_s$) was defined as the interval between the first and last times that $z_{cell}$ crossed 3 μm. Note that the mean $T_s$ remained relatively consistent even with variation in these two height criteria[14].

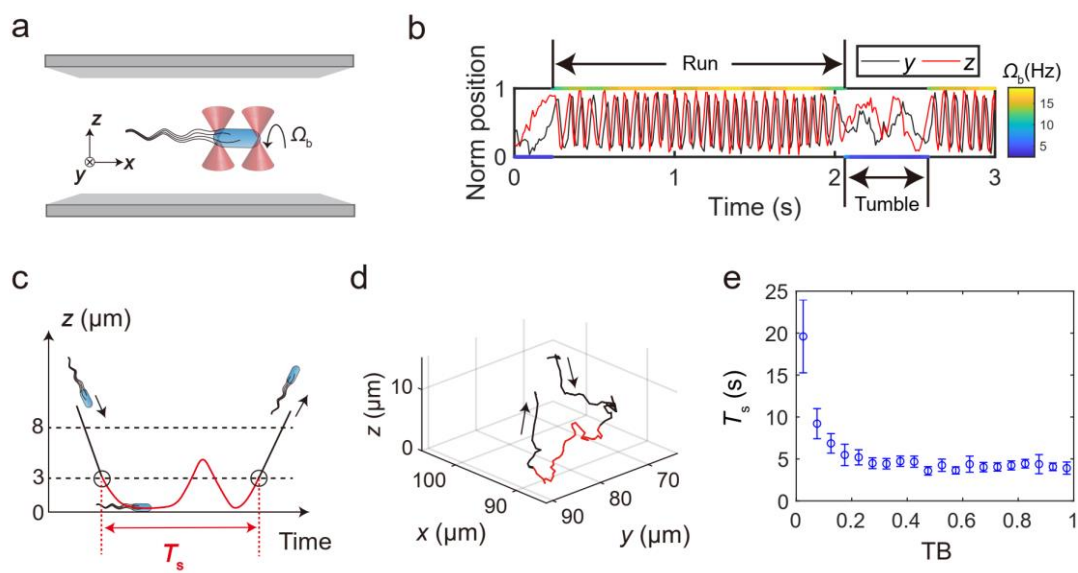

**Figure 1.** Surface residence times in relation to tumble biases. a-b) Illustration of run-and-tumble determination using optical trap signals. a) The bacterium is first trapped ~ 70 μm away from surfaces (not drawn to scale) to avoid noise from collisions or friction between the cell body and surfaces. The motion of cell poles is recorded as trap signals. b) The cell body's wobble (or rotating) frequency $\Omega_b$ (color bar) is derived from trap signals and used to differentiate between runs and tumbles. The position coordinates of cell poles are normalized in a 10-second time period. c) Schematic diagram defining surface residence time ($T_s$). After the trap signal measurement, the same trapped bacterium is moved near the surface and released from the optical tweezers. The subsequent bacterial path is represented by the solid curve, with the $T_s$ period highlighted in red. Two dashed lines indicate height criteria. d) An experimental bacterial near-surface trajectory after analysis and reconstruction. Black arrows indicate bacterial motion direction in c and d. e) Mean $T_s$ versus TB (bin size = 0.05). The total number of $T_s$ included in the statistics is 785 (from 98 biological replicates). Error bars denote SEM.

To reconstruct the three-dimensional trajectories of individual *E. coli* cells from the two-dimensional bright-field projections near the surface (Figure 1d), we developed an algorithm for extracting $z_{cell}$ using the halo width based on Rayleigh-Sommerfeld back-propagation[22,23] (see Supplementary Note S2, Figures. S2 and S3, Movies S2 and S3, Supporting Information). Considering that the medium temperature might vary due to absorption of the trapping laser energy—potentially altering refractive index of the medium, which was crucial for the tracking method—we monitored temperature variation during the trapping process using the temperature-sensitive fluorophore BCECF[24,25] (see Experimental Section). We found that when the trap laser was turned on, the temperature of the surrounding medium rapidly increased by ~ 1 °C within 10 seconds, then rose more slowly (Figure S4). The temperature variation was smaller at regions farther from the trap center. The maximum temperature increase during the entire trapping process (180 s) was ~ 1.5 °C. Conversely, when the trap laser was turned off, the temperature of the surrounding medium rapidly decreased by ~ 1 °C within 10 seconds, and then gradually decreased to near room temperature. Therefore, the temperature variation induced by the trapping laser (up to 1.5 °C) should not notably affect the refractive index (*n*) of the medium (for water at 645 nm illumination, $n \sim 1.3312$ at 23 °C and $n \sim 1.3310$ at 24.5 °C) or Brownian motion. Furthermore, when the cell was released from the traps by turning off the trapping laser, the rapid temperature decrease ensured that subsequent bacterial motion was recorded at near room temperature for 3D tracking.

We systematically analyzed hundreds of surface arrival-escape events involving bacteria with various run-and-tumble behaviors, and obtained $T_s$ and TB for individual events. As shown in Figure 1e, there is a rapid decrease in the mean $T_s$ as TB increases from 0, followed by a plateau phase transitioning around TB of approximately 0.25. Note that at very low TB (< 0.05), some bacterial trajectories exhibit $T_s$ values exceeding the maximum acquisition time of the camera (~ 207 s). In these cases, we were unable to observe the escape from the surface to calculate $T_s$, leading to an underestimation of the average $T_s$ in the first bin.

## 2.2 Phenomenological model for bacterial near-surface motion reproduces the experimental results

In previous motion models for tumbling *E. coli*, the bacterial center of mass is assumed to stay stationary without translational displacement, while the cell's orientation keeps changing until the next run event[10,14,26-28]. However, our observations reveal that tumbling cells (i.e., TB ~ 1) can rapidly escape from the surface (the last bin in Figure 1e). Consequently, the stationary-tumble assumption is not suitable for precisely modeling bacterial near-surface motion.

Considering the isotropic displacement of tumbling cells in the absence of spatial constraints[10], we assumed that the bacterial center of mass could be approximated as following translational Brownian motion. To test this hypothesis, we recorded the near-surface motion of *E. coli* cells locked in the tumble state (TB ~ 1), achieved by overexpressing CheY$^{13DK106YW}$. The mean square displacement in the *xy*-plane (MSD$_{xy}$), which is not constrained by the surface, is plotted against time (*t*), revealing a linear relationship (Figure 2a). Therefore, the translational motion of tumbling *E. coli* may be phenomenologically approximated as Brownian motion, which can be fitted with:

$$\text{MSD}_{xy} = 4D_t t, \tag{1}$$

where $D_t$ represents the effective translational diffusion coefficient in tumble states, with a fitted value of $6.61 \pm 0.03$ μm$^2$/s (Mean ± SD).

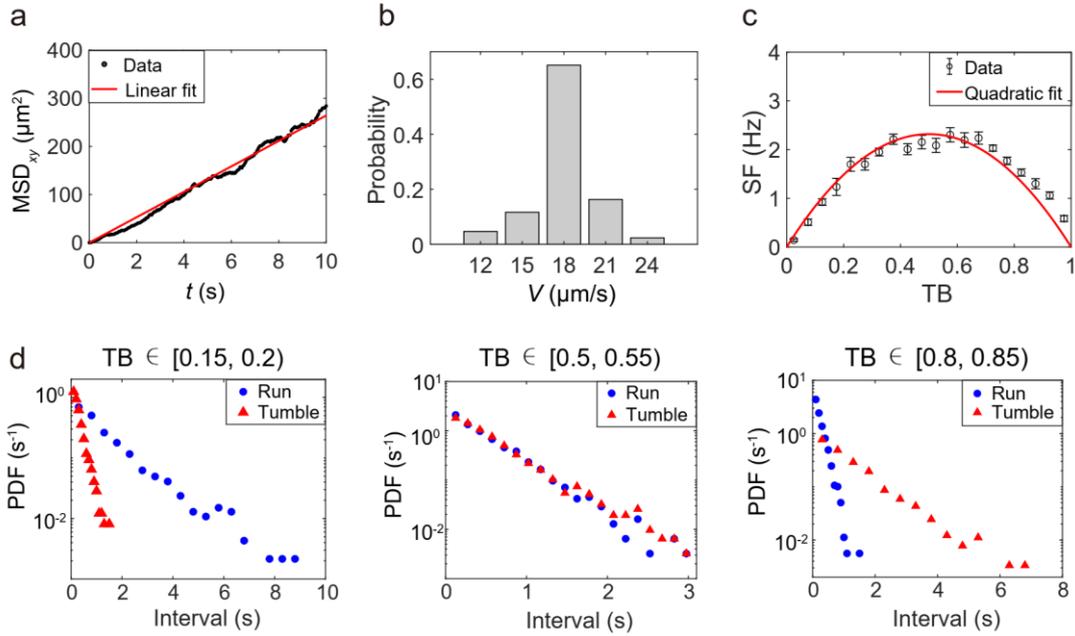

**Figure 2.** Analysis of run-and-tumble behaviors. a) MSD$_{xy}$ of 30 tumbling *E. coli* cells over time. The red line represents the result of linear fitting. b) Distribution of the swimming velocity (*V*) near the surface for 43 individual running *E. coli* cells. c) Mean SF versus TB for 246 individual *E. coli* cells. Error bars denote SEM. The red solid curve depicts the result of quadratic fitting as described by Equation (2). d) Distributions of run and tumble intervals after classification by TB (using the same sample as c). Experimental results in three different bins of TB are shown here as examples. Error bars denote standard deviation. PDF: probability density function.

To interpret the relationship presented in Figure 1e, we recorded near-surface swimming trajectories of a non-tumbling *E. coli* strain HCB1736 to determine the bacterial swimming velocity (*V*) near the surface (Figure 2b), which is $19 \pm 2$ μm/s (Mean ± SD). These non-tumbling bacteria seldom escaped from the surface within our observation timespan, consistent with a prior study[14]. Furthermore, we tried to understand the rules governing the transition between run and tumble states. Through analysis of run-and-tumble behavior sequences for individual *E. coli* cells, we determined two key metrics: TB and the switching frequency (SF) between run and tumble states. As shown in Figure 2c, the average SF is plotted against the corresponding TB. Given the symmetrical nature of this relationship, we applied a phenomenological quadratic fit:

$$\text{SF} = \frac{\omega}{4} - \omega(\text{TB} - 0.5)^2, \qquad (2)$$

where $\omega$ represents a characteristic frequency, with a fitted value of 9.27 ± 0.24 s$^{-1}$ (Mean ± SD).

Previous three-dimensional tracking studies indicated that both run and tumble intervals of wild-type *E. coli* cells follow exponential distributions[10,29,30]. Here, we also observed exponential distribution patterns in run and tumble intervals after categorizing them according to TB (Figure 2d). Therefore, transitions between run and tumble states can be treated as Poisson processes. In a steady state, the transition rates from run to tumble ($k_{RT}$) and from tumble to run ($k_{TR}$) can be expressed as:

$$k_{RT} = \frac{\text{SF}}{2(1 - \text{TB})}, \qquad k_{TR} = \frac{\text{SF}}{2\text{TB}}. \qquad (3)$$

In addition to the above measurements, previous studies found that changes in swimming direction during runs or tumbles can be represented as rotational diffusion[10,31], with the rotational diffusion coefficient during tumbles far exceeding that during runs[10,28,31]. Consequently, we proposed a phenomenological model for *E. coli*'s motion near surfaces (Figure 3a): During runs, a bacterium swims forward at a velocity $V$, changing its swimming direction with a rotational diffusion coefficient $D_r$. As illustrated in Figure 3b, before reaching the surface, the bacterial spatial motion can be described by stochastic differential equations (SDEs):

$$\frac{d\vec{r}(t)}{dt} = V\hat{e}(t), \qquad \frac{d\hat{e}(t)}{dt} = \sqrt{2D_r}\eta(t), \qquad (4)$$

where $\eta(t)$ is a Gaussian white noise of unit variance with $\langle \eta(t_1)\eta(t_2) \rangle = \delta(t_1 - t_2)$. To simplify the complex mechanisms of surface entrapment, we applied straightforward alignment rules[14], that is, when a running bacterium reaches the surface from the bulk, its orientation immediately aligns with the surface plane, as alignment time is negligible compared to typical surface residence time[6]. Subsequently, the bacterium swims along the surface in clockwise circular trajectories with a radius of curvature $R$ until the next tumble event[4,12], due to the long-range entrapment within our observation timescale. The boundary condition specifies that once a bacterium reaches the surface, $\varphi \equiv \pi/2$ until the next tumble. The corresponding SDEs are:

$$\frac{d\vec{r}(t)}{dt} = V\hat{e}(t), \qquad \frac{d\theta(t)}{dt} = \Omega + \sqrt{2D_r}\eta(t), \qquad (5)$$

where $\Omega$ is the angular speed approximated by $V/R$. During tumbles, the bacterium undergoes random walks resembling translational Brownian motion with diffusion coefficient $D_t$, and changes its orientation with rotational diffusion coefficient $D_\theta$. The corresponding SDEs are Langevin equations for pure diffusion:

$$\frac{d\vec{r}(t)}{dt} = \sqrt{2D_t}\xi(t), \qquad \frac{d\hat{e}(t)}{dt} = \sqrt{2D_\theta}\eta(t), \qquad (6)$$

where $\xi(t) = [\xi_1(t), \xi_2(t), \xi_3(t)]$, and $\xi_i(t)$ for $i$ = 1, 2, 3 are independent Gaussian white noises of unit variance with $\langle \xi_i(t_1)\xi_i(t_2)\rangle = \delta(t_1 - t_2)$. The displacement is constrained by the steric hindrance of the surface, meaning $z \geq 0$ as the boundary condition.

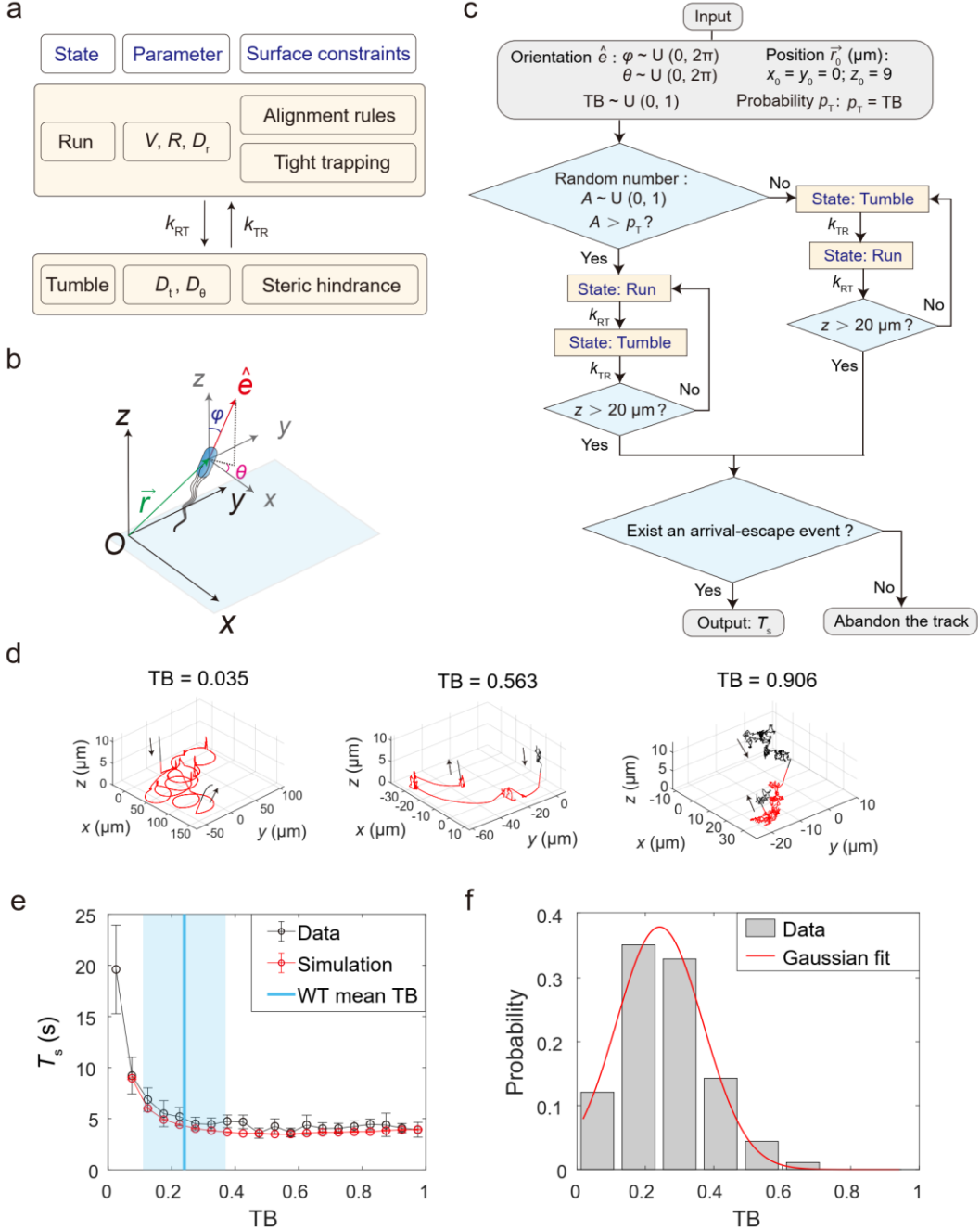

**Figure 3.** Modeling and simulation. a) Phenomenological motion model near surfaces, described as a simple two-state Markov chain. b) Illustration of the cell's centroid position $\vec{r} = (x, y, z)$, and the unit vector $\hat{e}$ indicating bacterial orientation with polar angle $\varphi$ and azimuth angle $\theta$. The surface is parallel to the $xy$-plane. c) Flow chart of simulating bacterial trajectories near the surface. d) Example simulated bacterial trajectories with various tumble biases. The radius of curvature $R$ for a run at the surface was set to 25 μm. Red segment of the trajectory

curve represents the $T_s$ period. Black arrows indicate the bacterial motion direction. e) Comparison between experimentally measured and simulated results of $T_s$ in relation to TB. Experimental results (black data) are copies of those in Fig. 1e. Simulated results (red data) are based on 100000 simulated trajectories. Error bars denote SEM. Blue line and shaded area indicate the mean and standard deviation of TB for wild-type *E. coli* cells, respectively. f) Distribution of TB for wild-type *E. coli* (91 cells). The Gaussian fit (red solid curve) peaks at 0.24 ± 0.13 (Mean ± SD).

To test the validity of the proposed model, we performed simulation to generate numerous bacterial near-surface trajectories using the proposed model (Figures. 3c-d, see Supplemental Note S3, Supporting Information), with all parameters set to experimentally measured values obtained from either this work or previous studies (Table S1, Supporting Information). The relationship between $T_s$ and TB was computed using 100000 simulated surface arrival-escape events with various TB values. For TB < 0.05, the mean $T_s$ exhibited significant fluctuations across multiple simulations (Figure S5, Supporting Information), likely because $T_s$ approaches infinity as TB approaches zero in the model, leading to variations caused by sampling error. This observation aligns with our experimental results, where measured $T_s$ for TB < 0.05 is also lower than the true value (the first bin in Figure 1e). For TB > 0.05, the mean $T_s$ in each of the bins remains relatively stable across repeated simulations, consistent with the experimentally measured results (Figure 3e). Furthermore, the simulated distribution of $T_s$ in each of the TB bins also agrees well with that of experiments (Figure S6, Supporting Information).

Using the dual-optical trapping method, we further measured the distribution of TB for wild-type (WT) *E. coli* cells, which peaks at ~ 0.24 (Figure 3f), in good agreement with the estimation from the mean CW bias of a single flagellum (Supplemental Note S1, Supporting Information). As shown in Figure 3e, the mean TB of WT *E. coli* cells is positioned close to the transition point between the dropping and the stable regions.

## 2.3 Bacterial surface diffusivity in relation to tumble bias

Microswimmers performing run-and-tumble motion have been shown to develop a diffusive motion in the long term[10]. A previous study demonstrated that enterohaemorrhagic *E. coli* (EHEC), which exhibits a run-spin-stop two-dimensional motility pattern distinct from conventional 3D run-and-tumble mode, achieves optimal transport on solid surfaces, highlighting the importance of efficient surface exploration for bacteria[32]. In our analysis, we consider the motion of planktonic *E. coli* cells as quasi-two-dimensional during surface residence time periods, which led us to examine bacterial surface diffusivity in relation to various run-and-tumble behaviors.

To investigate this relationship, we computed $MSD_{xy}$ during the surface residence time periods using numerous simulated bacterial trajectories, categorized based on their corresponding tumble biases (~5000 trajectories in each bin). To exclude the initial relaxation process for small time scale and determine the linear dependence of $MSD_{xy}$ on time $t$, we plotted the running diffusion constant (RDC), $MSD_{xy}/4t$, against time $t$. For this RDC-$t$ curve, the linear part of the $MSD_{xy}$-$4t$ relationship should correspond to a horizontal line with some fluctuation. As shown in Figure 4a, the boundary of phase I (relaxation process) and phase II (fitting time period) was determined by identifying where the RDC-$t$ curve begins to approach and converge to a horizontal line. For longer correlation time, the RDC-$t$ curve displays drastic fluctuation and deviates from the convergent horizontal line because of insufficient sample of long tracks, by which the boundary of phase II and phase III (insufficient data) was determined. Subsequently, $MSD_{xy}$ was plotted against $t$ to fit for the effective two-dimensional diffusion coefficient ($D_{xy}$) in the $xy$-plane using $MSD_{xy} = 4D_{xy}t$ (Figure 4b). For each of the TB bins, the fitting time period exceeds twice the length of mean surface residence time, and the statistical results exhibit similar features to those shown in Figures. 4a-b. Within the physiological ranges of parameter $R$ (15-35 μm) and $V$ (19-25 μm/s)[4,10], the relationship between $D_{xy}$ and TB consistently shows a peak near TB ~ 0.25 (see scattered markers in Figure 4c, and Figure S7, Supporting Information). Notably, we found that the mean TB (~ 0.24) of wild-type *E. coli* cells is situated near the peak position of the $D_{xy}$ versus TB relationship (Figure 4c).

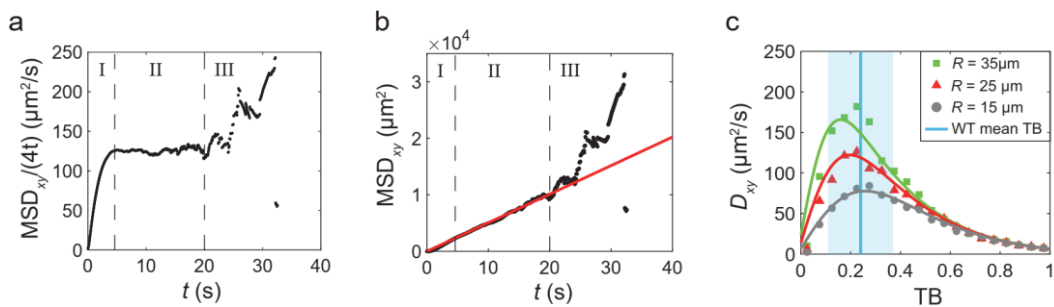

**Figure 4.** Bacterial diffusivity during the surface residence time periods. a-b) Computation of bacterial diffusivity during the surface residence time periods using simulated bacterial trajectories. Statistical results in the TB range of (0.2, 0.25) are presented as illustrative examples ($R$ = 25 μm). a) $MSD_{xy}/4t$ versus time $t$. Areas with different labels represent: I, relaxation process; II, fitting time period; III, insufficient data. b) $MSD_{xy}$ versus time $t$. Red line represents linear fit to the data within the fitting time period. c) $D_{xy}$ versus TB with different values of the parameter $R$. Scattered markers denote $D_{xy}$ values from simulation, and solid curves represent the relationships computed with Equation (12), where the radius of curvature $R$ was adjusted to $C \times R$ for the computation ($C$ is a correction factor with values of 1.2, 1.3 and 1.5 for $R$ = 35, 25 and 15 μm, respectively). Blue line and shaded area indicate the mean and standard deviation of TB for wild-type *E. coli* cells, respectively.

## 2.4 Theoretical analysis of bacterial surface diffusivity

To better understand the relationship shown in Figure 4c, we further carried out theoretical analysis to obtain a numerical solution. In our model, bacterial surface behavior can be described by a simple two-state Markov chain consists of run and tumble states (Figure 3a), with the master equations:

$$\frac{\partial p_R(t)}{\partial t} = -k_{RT} p_R(t) + k_{TR} p_T(t),$$

$$\frac{\partial p_T(t)}{\partial t} = -k_{TR} p_T(t) + k_{RT} p_R(t), \quad (7)$$

where $p_R(t)$ and $p_T(t)$ denote the probability of finding the bacterium in state run and tumble at time $t$, respectively. We only consider the bacterial motion in the $xy$-plane, consequently, the probability density of finding the bacterium in state R (run) or T (tumble) at position $\vec{r} = (x, y)$ with azimuth angle $\theta$ at time $t$, denoted as $p_{k=R,T}(\vec{r}, \theta, t)$, can be obtained from the spatiotemporal evolution described by the following Fokker-Planck equations with state transition jumps:

$$\frac{\partial p_R(\vec{r}, \theta, t)}{\partial t} = -V\hat{e}(t) \cdot \nabla p_R(\vec{r}, \theta, t) - \Omega \frac{\partial p_R(\vec{r}, \theta, t)}{\partial \theta} + D_r \frac{\partial^2 p_R(\vec{r}, \theta, t)}{\partial \theta^2}$$
$$- k_{RT} p_R(\vec{r}, \theta, t) + k_{TR} p_T(\vec{r}, \theta, t),$$

$$\frac{\partial p_T(\vec{r}, \theta, t)}{\partial t} = D_t \nabla^2 p_T(\vec{r}, \theta, t) + D_\theta \frac{\partial^2 p_T(\vec{r}, \theta, t)}{\partial \theta^2}$$
$$- k_{TR} p_T(\vec{r}, \theta, t) + k_{RT} p_R(\vec{r}, \theta, t). \quad (8)$$

Note that the experimentally extracted three-dimensional values of $D_r$ and $D_\theta$ can be used to approximate the two-dimensional rotational diffusion coefficients, because the direction correlation of spatial reorientation is comparable to that of its projection in the surface plane[31]. To solve the equations, the initial conditions can be concluded as: i) The probability densities of finding the bacterium in state R (run) and T (tumble) at $t = 0$ are 1-TB and TB, respectively, that is,

$$\int d\vec{r} \int d\theta \, p_R(\vec{r}, \theta, t = 0) = 1 - \text{TB},$$

$$\int d\vec{r} \int d\theta \, p_T(\vec{r}, \theta, t = 0) = \text{TB}. \quad (9)$$

ii) The bacterial initial position ($\vec{r}|_{t=0}$) is randomly distributed in the surface ($xy$-plane). iii) The initial azimuth angle ($\theta|_{t=0}$) is randomly distributed in the range [0, 2π). The boundary condition can be concluded as the bacteria keep moving in an infinite $xy$-plane.

The spatial dynamics of a self-propelled particle follows:

$$\frac{d\vec{r}(t)}{dt} = \vec{v}(t), \quad (10)$$

where $\vec{v}(t)$ denotes the two-dimensional velocity of the particle at time $t$. The mean squared displacement (MSD$_{xy}$) is directly related to the velocity correlation function, by direct integration of Equation (10) and using the symmetry of the velocity correlation function with respect to permutation of the times $t'$ and $t''$, known as the Taylor-Kubo formula[33]:

$$\text{MSD}_{xy}(t) = \langle |\vec{r}(t) - \vec{r}(0)|^2 \rangle = 2 \int_0^t dt' \int_0^{t'} dt'' \, \langle \vec{v}(t') \cdot \vec{v}(t'') \rangle. \quad (11)$$

By solving the Fokker-Planck equations and using Equation (11) to simplify the computation of diffusion coefficient (Supplementary Note 4, Supporting Information), we obtained the analytical form of $D_{xy}$:

$$D_{xy} = \text{TB} D_\text{t} + \frac{V^2}{2}(1 - \text{TB}) \frac{(D_\theta + k_\text{TR})(D_\text{r} k_\text{TR} + D_\theta k_\text{RT} + D_\text{r} D_\theta)}{(D_\theta + k_\text{TR})^2 \frac{V^2}{R^2} + (D_\text{r} k_\text{TR} + D_\theta k_\text{RT} + D_\text{r} D_\theta)^2}. \quad (12)$$

The two terms in the sum of Eq. 12 represent the contributions from the tumble and run periods, respectively. As $R$ increases from 15 to 35 μm, the local peak of $D_{xy}$ calculated by Equation (12) shifts moderately to the left, with the peak position close to the mean tumble bias of wild-type *E. coli* (Figure S8, Supporting Information). The directly calculated values of $D_{xy}$ by Equation (12) are slightly lower than those from simulation (Figure 4c, scattered markers), mainly because within the surface residence time, the bacterium swims in circular trajectories for most of the time in run periods. However, it will occasionally swim in straight paths when $z_\text{cell} > 0$ in the simulation. This effect can be roughly approximated as an increase of the mean radius of curvature $R$ (by multiplying $R$ by a correction factor $C > 1$) in Equation (12) (Figure 4c, solid curves). The general computed shapes of the $D_{xy}$ versus tumble bias relationship coincide with those from the simulation, both demonstrating the optimal surface diffusivity of wild-type *E. coli*.

## 3. Conclusion and Discussion

Bacteria tend to accumulate at significantly higher concentrations on solid surfaces than in bulk liquid[5-8,12,19]. However, several physical constraints imposed by surfaces greatly limit the efficiency of bacterial environmental exploration[4,6-8]. In this study, we focused on surface residence time, which serves as an indicator of surface escape ability, in relation to TB for individual *E. coli* cells. As TB increases from 0, the mean surface residence time initially decreases sharply and then stabilizes at a minimum value of approximately 4 s when TB exceeds 0.25. We quantitatively characterized run-and-tumble behaviors in relation to bacterial TB and developed a phenomenological motion model for *E. coli* cells swimming near surfaces by simplifying the complex mechanisms of physical interactions.

A key difference between our model and previously proposed ones lies in the translational motion of tumbling cells. We found that tumbling cells are able to rapidly escape from the surface, thus rejecting the stationary assumption for their centers of mass widely used in earlier models[10,14,26-28]. During tumbles, some clockwise-rotating flagella break free from the flagellar bundle. Concurrently, the bacterium experiences thrust from multiple directions due to the randomly distributed flagella around the cell body. Additionally, the flexibility of the flagellar hook, which connects the motor and filament[34], further randomizes the thrust. The cell axis also undergoes random reorientation because the torque exerted by the flagellar motors on the filaments must be balanced by counterrotation of the cell body[10]. Thus, a tumbling bacterium continuously changes its swimming direction randomly, and its translational displacement appears isotropic in the absence of spatial constraints[10], resembling Brownian motion.

However, the diffusive assumption for tumbling *E. coli* cells adopted here requires testing, since it implies that cells escape from the surface via diffusion. In reality, the bacterial escape mechanism likely involves complex interactions including hydrodynamic interaction, Brownian motion, flagella-surface collision, among others. This raises the question of whether the bacterial motion in the normal direction (*z*-direction) of surface plane can also be modeled as a diffusive process. Direct validation of this near-surface diffusive assumption in the *z*-direction remains challenging due to spatial constraints imposed by the surface. To address this, we performed simulations and found that analysis of simulated bacterial trajectories closely matched the experimentally measured relationship between surface residence time and TB, thereby validating the model on the timescale of our measurements.

Therefore, we conclude that the combined effect of various factors on the near-surface motion of tumbling *E. coli* can be phenomenologically approximated as translational Brownian motion with steric hindrance of the surface. We further measured the TB distribution of wild-type *E. coli* cells, which peaks at approximately 0.24, aligning with the transition point between the decreasing region and the plateau in the surface residence time versus TB relationship.

Utilizing this model, we generated a large quantity of simulated bacterial trajectories. We then analyzed the effective two-dimensional diffusion coefficient during the surface residence time periods, correlating it with TB for individual bacteria. Our findings reveal that the surface diffusivity peaks near the mean tumble bias of wild-type *E. coli* cells. Theoretical analysis of bacterial surface diffusivity captures the general shape of the $D_{xy}$ versus TB relationship observed in simulations, further supporting the existence of this optimal diffusivity.

For individual cells, the efficiency of surface exploration scales with the surface area covered per unit time, which in turn scales with the surface diffusivity $D_{xy}$. Considering the heterogeneity in the steady-state TB of wild-type *E. coli* cells, which is mainly induced by noise in gene expression, this reflects a strategy for bacterial offspring persistence under the physical constraints of their habitat[35]: Without stimulus cues, bacteria cannot determine whether solid surfaces contain livable sites. Under these conditions, some bacteria (with large TB) can escape from nearby surfaces in the shortest time to regain greater freedom of movement in bulk liquid if the surface lacks nutrients. Meanwhile, for those that spend more time on the surface (with smaller TB), they perform the most efficient two-dimensional exploration, which increases the chances of finding randomly distributed food patches.

This optimal surface exploration strategy also appears in other bacterial species, such as EHEC, which displays a unique run-spin-stop motion pattern at two-dimensional surfaces[32], distinct from the conventional run-and-tumble motion of planktonic *E. coli*[10]. These similar behaviors suggest that surface exploration efficiency may be a critical factor in bacterial evolution.

In conclusion, we reveal the significance of bacterial adjustment of TB in uniform environments. Bacteria exhibit various types of surface-associated motion, with swarming, twitching, and gliding frequently observed near surfaces[36]. They adapt their behaviors to different surface conditions, as seen in swarming *E. coli* cells, which forego tumbling and occasionally reverse their direction of motion[37]. Furthermore, contact with surfaces is crucial for biofilm formation[3]. Bacteria employ diverse motion patterns in response to their interactions with surfaces, suggesting that many survival mechanisms likely remain to be discovered in future studies.

## 4. Experimental Section

***Strains and plasmids***: In this study, two *E. coli* strains were used to perform the experiments. The strain referred to as 'wild-type' is HCB1 (AW405)[38], while the strain HCB1736 ($\Delta cheY$) is a derivative of HCB1. The plasmid pBAD33-*cheY*$^{13DK106YW}$ expresses the mutant CheY$^{13DK106YW}$ that is constitutively active even without phosphorylation[39] under control of an arabinose-inducible promoter[40]. As a result, the fluctuation of intracellular CheY-P concentration due to stochasticity in the chemotaxis network is absent[41,42], which leads to a stable tumble bias on the timescale of our measurements. To measure the run-and-tumble behaviors and near-surface motion of individual *E. coli* cells, HCB1736 transformed with pBAD33-*cheY*$^{13DK106YW}$ was utilized. Tumble-locked *E. coli* cells was achieved by overexpressing CheY$^{13DK106YW}$. HCB1736 was used to measure the smooth swimming velocity of *E. coli* cells near the surface. HCB1 was used to measure the distribution of tumble bias of wild-type cells.

***Cell culture and sample preparation***: For all strains, a single colony from an LB-agar plate was isolated and grown overnight in LB. The overnight culture was then diluted 100-fold into 10 ml of tryptone broth, and grown to mid-log phase (OD$_{600}$ = 0.5) at 33°C. For the strain HCB1736 transformed with pBAD33-*cheY*$^{13DK106YW}$, 25 μg/ml chloramphenicol and 0.001-0.005% arabinose were added to induce bacterial motility across the entire range of run-and-tumble behaviors, encompassing a tumble bias from 0 to 1. Subsequently, the cells were washed twice with trap motility buffer (TMB)[17,20], which contained 70 mM NaCl, 0.1 mM methionine, 100 mM Tris-Cl, 2% glucose, and an oxygen-scavenging system (80 μg/ml glucose oxidase and 13 μg/ml catalase). The oxygen-scavenging system effectively mitigates potential photodamage induced by the optical traps due to the high photon flux at near-infrared wavelengths[20,43,44]. Finally, an aliquot of washed cells was gently diluted 200-fold into TMB, and added into the sample chamber, which was constructed with a slide and a coverslip using 150-μm thick double-sided tapes as spacers. Both the glass slide and coverslip were pre-cleaned as described in our previous work[45]. The sample chamber size was ~ 20 mm × 20 mm × 150 μm (length × width × height).

***Optical traps and microscopy***: Experiments were performed using a commercially available optical trap instrument (NanoTracker 2, JPK Instruments). The dual-optical traps were constructed using a single 1064-nm diode-pumped solid-state laser, which was split into two orthogonally polarized beams. Both beams were tightly focused by a 60×, water-immersion (1.2 NA) microscope objective (Nikon) beneath the sample chamber to generate two separated optical traps. The separation between the two traps was controlled by a piezo-actuated mirror stage. The laser power has a maximum output of ~ 5W and can be adjusted using the laser control software provided by the manufacturer. In our experiments, we used 50 mW of trapping power at the sample plane per optical trap, which was sufficient to stably trap and manipulate the cells while minimizing photodamage. An identical objective lens placed above the sample chamber was used to collect transmitted light for position detection and bright-field imaging. The resolution of the axial variation in trap position was 1 nm. The lateral size of each trap in the *xy*-plane (normal to the trap beam, as shown in Figure 1a) was about 650 nm, and the axial size was about 1 μm. The principles of signal detection have been described previously[20]. Briefly, to determine runs and tumbles, the motion of the two cell poles, which were trapped by two separate optical tweezers, was detected by position-sensitive photodetectors. Bright-field illumination was provided by a condenser LED light source with a long-pass glass filter (Thorlabs, FGL645) to minimize cell damage from short-wave light[46]. The resolution of the bright field system can be estimated using the formula $0.61\lambda/NA$, yielding approximately 328 nm. Details of the 3D tracking near surface are presented in the Supplementary Information.

***Data acquisition***: For the measurement of run-and-tumble behaviors, individual *E. coli* cells with excellent motility were randomly selected without additional filtering. The trapped cell was moved a sufficient distance away from both the top and bottom surfaces (~ 70 μm) to avoid additional noise from collisions or friction between the cell body and surfaces. Raw data (motion of cell poles) for analyzing run-and-tumble behaviors were recorded for 3 min at a sampling frequency of 1,000 Hz. By this means, the bacterial tumble bias (TB) and run-and-tumble switching frequency (SF) can be analyzed from the whole run-and-tumble time sequence. Following this, the same trapped cell, situated 4 μm above the focal plane, was displaced about 9 μm away from the coverslip surface and further to a selected field of view with low cell density and few motile cells. This procedure minimized potential effects from collisions, hydrodynamic interactions, and overlapping tracks with other swimming cells, which could interfere with the analysis of cell-surface interactions. The cell was then released from the optical traps, and its motion near the surface was recorded at 25 fps (exposure time: 5 ms) under bright-field illumination with a CMOS camera (Phantom, Miro, LAB3a10) equipped with a 12 GB high-speed internal RAM, allowing for a maximum acquisition time of approximately 207 s in the full field of view (213 × 213 μm). The video recording was terminated either until the maximum acquisition time or when the bacterium was too far from the focal plane to be distinguished. Furthermore, if the cell escaped from the surface immediately following a collision with other cells, the data were discarded, as such escape events were likely caused by cell-cell interactions rather than intrinsic cell-surface interactions. Once released, the same bacterium would not be trapped again. Therefore, the data were randomly collected from a series of experiments using hundreds of bacteria with various amounts of added inducer (arabinose). Note that even under the same amount of added arabinose, bacterial tumble bias will display some variation for individual cells because of the phenotypic differences. The motorized sample stage was manually adjusted in the *xy*-plane to keep the target cell in view throughout the recording process.

***Run-and-tumble behavior analysis*:** The procedure for determining runs and tumbles from the optical trap signals has been thoroughly described elsewhere[17,20]. Briefly, during run periods, all the flagella rotate counterclockwise and form a single bundle, and cell poles display oscillatory wobble due to the counterrotation between the bundle and the cell body, whose axis is not collinear with the bundle[47]. During tumble periods, some of the flagella switch to clockwise rotation, consequently, the bundle flies apart, leading to erratic patterns of the motion of cell poles. The instantaneous body wobble frequency $\Omega_b$ can be derived from the position signals of cell poles by continuous wavelet transform and used to differentiate between run (high frequency) and tumble (low frequency) periods. By applying a predefined threshold for body-rotation (wobble) frequency, run and tumble periods could be distinguished. Note that in this case, the definitions of 'run' and 'tumble' are equivalent to flagellar 'bundling' and 'unbundling', respectively, as described by Min *et al.*[20]. TB was extracted as the fraction of time spent tumbling for each trace.

***Monitoring of temperature variation during the trap process*:** BCECF (2',7'-bis-(2-carboxyethyl)-5-(and-6)-carboxyfluorescein, B1151, ThermoFisher Scientific Inc., United States) was dissolved in 50 mM Tris-Cl buffer (pH 7.2) at a concentration of 1 mM. The calibration of normalized fluorescence-temperature relationship (Fig. S4a) of the BCECF solution was measured using a microscope equipped with a custom-designed constant temperature device (temperature accuracy: ± 0.1 °C), which includes an air heater, a vibration-free circulation fan, an acrylic glass enclosure, and a feedback-regulated temperature control system (see our previous study for details[25]). An experimental sample chamber filled with BCECF solution was used to monitor temperature variation during the trap process. The medium fluorescence was excited by a condenser LED light source with appropriate filter sets and recorded for 420 s. The optical trap laser was turned on at 60 s and turned off at 240 s (corresponding to a trapping time of 180 s in our data collection). The fluorescence variation was then converted to temperature variation (Fig. S4c) using the linear fluorescence-temperature calibration curve (Fig. S4a).

***Statistical Analysis:*** Data in each TB bin was shown as mean ± standard error of the mean (SEM). Sample sizes for each statistical analysis were indicated in the figure legends. The Gaussian distribution of TB of wild-type *E. coli* cells was first assessed using the Anderson-Darlin test, then the normalized probability in each bin was used for Gaussian fitting. The probability density functions (PDF) of run and tumble intervals in each TB bin were calculated using the normalized probabilities and bin sizes. All statistical analyses were performed using Matlab R2024a.

## Acknowledgements

This work was supported by National Natural Science Foundation of China Grants (11925406 and 12090053) and a grant from the Ministry of science and technology of China (2019YFA0709303).

## Conflict of Interest

The authors declare no conflict of interest.

## Author Contributions

J. Y., A. T. and R. Z. designed the work, A. T. and G. L. performed the experiments, analyzed the data and drew the figures. All authors wrote the manuscript. A. T. and G. L. contributed equally to this work.

## Data Availability Statement

The data that support the findings of this study are available from the corresponding author upon reasonable request.

## Keywords

microswimmers, bacterial motion, surface constrains, environmental exploration, bet-hedging strategy

## Supporting Information includes:

Notes S1 to S4
Table S1
Illustration of Movies S1 to S3
Figures S1 to S8